*Research Article*

# Enhancing Safety in Water Transport System Based on Internet of Things for Developing Countries

**Md Mohaimenuzzaman,[1] S. M. Monzurur Rahman,[1] Musaed Alhussein,[2] Ghulam Muhammad,[2] and Khondaker Abdullah Al Mamun[1]**

[1]*Advanced Intelligent Multidisciplinary Systems Lab (AIMS Lab), Department of Computer Science and Engineering, United International University, Dhaka 1209, Bangladesh*
[2]*Department of Computer Engineering, College of Computer and Information Sciences, King Saud University, Riyadh Riyadh 11543, Saudi Arabia*

Correspondence should be addressed to Md Mohaimenuzzaman; mohinr@gmail.com

Accidents in inland waterways in developing countries are a regular phenomenon throughout the year causing deaths, injuries, monetary loss, and a significant amount of missing people. In consequence, a lot of families are losing their dear ones leading to much misery. The above context demands an intelligent, safe, and reliable water transport system for the developing countries. The concept of Intelligent Transport System (ITS) can be applied to develop such system; however, there are issues with ITS and Internet of Things (IoT) unlocks a new way of developing it. This paper proposes a model to transform the water transport system into an intelligent system based on IoT. IPv6 based machine-to-machine (M2M) protocol, 3G telecommunication technology, and IEEE 802.15.4 network standard play a significant role in this proposed IoT based system.

## 1. Introduction

The waterways play a significant role in the transportation system of many developing countries because of their lower cost and higher accessibility compared to other alternatives, creating a great demand for transportation of goods and passengers. However, this mode of transportation has become vulnerable due to the limitation of resources, lack of care, and the absence of technology to maintain and monitor the waterways and the water-based vehicles. The scenario is pretty much the same for most of the developing countries [1]. Due to the availability of information and ease of access to relevant information, we choose Bangladesh as a case study for this research.

Water transportation system is a very important mode of transportation for the people of Bangladesh, especially in the southern part of the country [2]. Moreover, in some areas of the country, water transport is the only mode of public transport available. According to Bangladesh Internal Water Transport Authority (BIWTA) about 7% (about 24,000 km) of the country's surface is covered by inland waterways. Over eighty-seven million passengers are carried through this system annually [3].

As the population of Bangladesh is becoming denser, the waterways are also becoming congested [4]. Accidents have become a regular phenomenon in this sector. Awal showed in his research that several thousand people have met premature death or injuries or have been reported missing due to accidents in the waterways in the past few years [5]. His further investigation revealed that the number of accidents has increased significantly over the years, and the most predominant reasons behind the increase were passenger overloading and storms including cyclones and collisions. On February 22, 2015, a passenger launch was hit by a cargo vessel in the middle of the mighty Padma River and the death toll was over 70, and few went missing [6]. A few months back, on August 4, 2014, around 11:00 AM a launch sank in choppy waters and strong winds in the same river, but at a different location, and over 100 passengers lost their lives, and 87 people went missing. The launch was also lost as the rescuers were unable to find it in the deep water [7]. Besides loss of lives, the loss incurred through damage to properties puts an enormous amount of burden on the national economy. Although the government has taken some active measures in response to such emergencies, the safety scenario has improved by a negligible amount [8]. The most unfortunate aspect is that the rescue operations take too long to get initiated due to a lack of efficient communication infrastructure, which increases the damage manifolds.

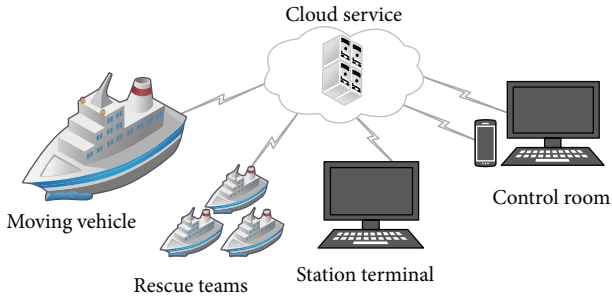

Figure 1: Overall architecture of Intelligent Water Transportation System (IWTS).

Bangladesh, along with other developing countries, is also facing difficulties in providing secure and effective transportation systems to its people in roadways. Intelligent Transport System (ITS) is widely being implemented to make the urban transportation system safe and efficient [9, 10]. Due to lack of infrastructure and other limitations of the road transportation system of the developing countries, tremendous effort and large investment will be needed to establish an intelligent system for roadways. On the other hand, an investment in an intelligent system for internal waterways would be more feasible as only the vehicles are needed to be incorporated into the intelligent system, not the waterways.

Upon considering the above facts and the importance of the water transportation system for the developing countries like Bangladesh, this paper focuses on developing a new transport model based on the Internet of Things (IoT) that transforms the unsafe waterways to a safer, more reliable, and sustainable system. The proposed model is identified as Intelligent Water Transportation System (IWTS) for developing countries, and Figure 1 provides an overall architecture of it. According to IWTS, different monitoring devices comprised of machine-to-machine (M2M) system technologies are installed in the water-based vehicles. The monitoring data are transmitted from the vehicle to the cloud service in definite intervals for analyzing and detecting emergency. As the implementation detail of the complex data processing and emergency prediction system of the cloud service is itself a candidate of a separate research topic, it is considered to be beyond the scope of this paper. However, various proposals from different researchers and the technical details of the proposed model are covered in the forthcoming section. This paper is organized as follows: Section 2 summarizes the study on the related technologies and works based on those technologies prior to developing IWTS. Section 3 describes the proposed model in detail. Section 4 describes the effects of IWTS in the safety measures of waterways, Section 5 shows the experimental results that support the claims this paper does in Section 4, Section 6 shows some open area for further research, and, finally, Section 7 concludes the paper.

## 2. Background Study

According to the discussion in Section 1, the predominant reasons for mishaps in waterways are overloading, storms, cyclones, and collisions. In order to ensure safety, the vehicles need to stop overloading passengers, be as safe as possible from natural calamities, collisions, and be taken for regular maintenance. Thus the vehicle in water needs to have a system comprised of different information sensing devices to determine the number of passengers on board, the total weight of passengers and freight, location, and speed, monitor its freeboard (the height of the water-based vehicle's deck above the water level), the tidal wave height, and water current speed and direction, and monitor the weather condition and so forth to avoid devastating situations. The system also needs to send the sensed data to the application service for processing and storing, where processing of data uses techniques of big data processing and data mining [9]. The cloud service needs to analyze the vehicle situation and send notification about issues to the vehicle. In case of the vehicle safety alerts, it also needs to make communication between the vehicle, the transport authority, and potential rescue teams, and the station terminal in real time. Hence, this problem looks like a similar problem that can be solved using the concept of ITS; however, IoT provides new dimensions and strong potentials that can be used to enhance the efficiency of ITS [10, 11].

The IoT is an environment, where different physical objects or "things" are embedded with software programs, sensors, and electronics and are provided with unique identifiers as well as the ability to collect and exchange data over a network without requiring any human interaction [9, 12]. Hence, the physical objects can be featured as tiny computers referred to as smart things [12]. According to the definition of IoT, the smart objects or things do their work without human interaction which means they work in a machine-to-machine way and this is known as the M2M technology, an integral part of IoT. Devices comprising M2M, referred to as smart devices, are equipped with a wide range of environment and context sensors for collecting data, as well as with multiple personal, local, and wide area communication capabilities [13, 14]. M2M applications are basically used for the purposes of monitoring (e.g., environment, weather, and temperature), controlling devices and tracking things [13]. Nowadays, M2M gets the preference over Wireless Sensor Network (WSN) because M2M not only enables the devices to deliver sensor-related data but also enables autonomous communications between M2M devices without human interactions [15]. Thus, every M2M device needs to have a unique IP address to avoid isolation from other networks, which presents a new challenge as IPv4 is unable to provide such a vast amount of addresses. IPv6 is the only alternative in this case as it provides that massive amount of unique addresses along with some attractive features like stateless configuration and small sized easily manageable header [16].

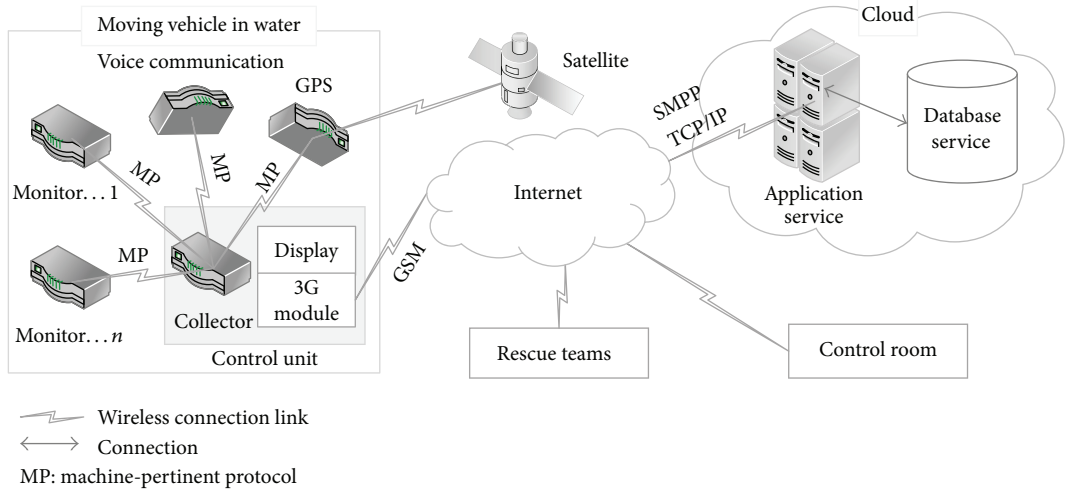

Figure 2: Detailed architectural view of IWTS model.

However, applications using these technologies usually have vast networks that connect a lot of interoperable smart devices and applications. Thus, less power consumption is of utmost importance to ensure longer network lifetime. IEEE 802.15.4 ZigBee protocol was specially developed for wireless devices ensuring that it is long lasting and requires low power [17, 18]. A comparative study by Lee et al. reveals that the ZigBee technology consumes lower power for both transmission and reception compared to that by the Wi-Fi and UWB technologies [19]. To ensure wireless communication facility among the devices in such networks, 3G modules are used because they support multiple frequency bands and integrate network application protocols (e.g., MMS, TCP, IP, SMTP, and FTP). 3G module like SIM5216E supports HSDPA/WCDMA/GSM/GPRS/EDGE, which ensures the availability of wireless communication in case of unavailability of 3G network coverage [20].

There are related works using the above technologies by different researchers. Yongjun et al. proposed an improved and efficient ITS for urban traffic based on IoT [21]. Cardoso et al. defined the feasibility of using IoT architecture for more complex control and traffic monitoring using Radio Frequency Identification (RFID), Low Level Reader Protocol and embedded device built in using M2M technology [9]. Huang and Liu proposed a more detailed oriented urban transportation management system based on IoT [10]. They described the road map of updating the existing ITS based on the integration of IoT and embedded devices using M2M technology. They also proposed a concept of cloud computing that played a major role in their proposed system. Although there are urgencies of ITS for waterways for the developing countries, we see the proposed systems are applicable for roadways only. There is no such system for waterways to be realized yet even with the technologies that were used in ITS before the concept of IoT came into reality. This paper analyzes the situation in detail and proposes such an ITS for the waterways to ensure safety for the water-based vehicles.

## 3. Intelligent Water Transportation System (IWTS)

According to the proposed model, the moving vehicle has a control unit and a number of M2M devices installed. The control unit is comprised of a special M2M device called the collector, a display unit for displaying information, and a 3G module for wireless communication. The collector is connected with all other M2M devices, the GPS system, and a voice communication system. The collector translates the signals transmitted from the devices into meaningful data, puts all data together in a specific format, and uses the wireless communication system provided by the installed 3G module; it also sends the data to the cloud service. The cloud service analyzes the data with the help of the trained decision model to predict the present situation of the moving vehicle in the water. In case of any emergency predicted by the cloud service, it sends an alert message with detailed information to the control room, the nearest rescue teams, and station terminal and to the vehicle using GPRS system. Finally, it stores the received data and other related information to the database. Figure 2 shows the detailed architectural view of the proposed model.

The technical detail of the proposed model is described below.

*3.1. Monitors.* Monitors are task specific, sensor based, smart devices comprising wireless M2M technology for information sensing. To monitor the weight taken on board, the wave height, water current speed and direction, freeboard (the height of the water-based vehicle's deck above the water level), and weather condition, M2M communication technology is used. According to Lawton, this type of communication technology is best used for such monitoring as it can operate autonomously and push information to other devices as well as being able to make some decisions on its own [13]. Furthermore, M2M devices are low powered devices, and they are not used to take complex decisions. To determine

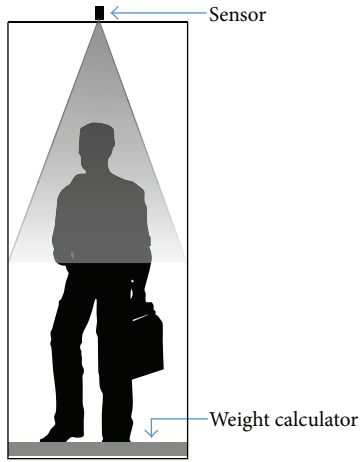

Figure 3: Monitor for passenger counting and weight calculation.

the location of the vehicle, time, and speed, the GPS tracking system is used [22].

The passenger counting monitor has an autonomous directional sensor (e.g., DA-200 directional sensor) to keep track of passengers getting on and off the vehicle [23]. It is also comprised of a weight calculating device to make it able to determine the passenger weight as well. This monitor is installed in the entrance and exit doors and only transmits data when a passenger gets on or off the vehicle. Figure 3 shows such a monitor installed in the entrances and exits of the water-based vehicle. In technologically advanced countries, RFID readers are used to count the number of passengers getting on and off a vehicle [21]; however, it might be an ambitious task right now for a developing country like Bangladesh as every passenger needs to have a card with an RFID tag.

Monitors except the passenger counter are of similar types, but each one is responsible for a specific task. They periodically transmit messages to the collector. However, if they keep transmitting messages for the collector in a regular interval, the collector has to receive a high number of messages from multiple monitoring devices. As the M2M devices have limited processing capability and low memory, they need to transmit messages in a more efficient way. One solution could be of transmitting a message only when necessary. Let us assume that a monitor denoted by $M_i$ transmits messages in an interval $\Delta$ time. If the message to be transmitted by $M_i$ remains the same at $\Delta_t$ and $\Delta_{t+1}$ time, $M_i$ should not be transmitting the message at time $\Delta_{t+1}$; rather it should stay silent until a different message needs to be sent. By incorporating this intelligence level into all the monitors, it can be ensured that the collector is not going to be overwhelmed by incoming messages. Thus, it also ensures low processing power requirements for the monitoring devices. Since the M2M based monitors can take some decisions on their own, they can determine a situation that is a potential candidate for emergency. In such cases, they send the message to the collector immediately regardless of the next interval of sending messages.

### 3.2. Control Unit

*3.2.1. Collector.* The collector is a special purpose M2M device that listens to all the monitors, the GPS system, and the voice communication system. Matamoros and Anton-Haro and Boswarthick et al. used such a device where a group of M2M devices act as monitors and a separate M2M device gathers information by reading the transmitted signals from the monitors and sends the information to the application service using its wireless communication system [24, 25]. In IWTS, collector stores the monitoring data in its memory for further use as the monitors stay silent until a fresh message needs to be sent. The collector has a timer and when it expires, the collector initiates a new timer and sends a request to the silent monitors for messages. Moreover, if an emergency category message is received from any monitor, the collector immediately requests other monitors to transmit messages. After receiving the messages from different monitors and the GPS system, it interprets them and combines the data to a specific format (e.g., JSON) and sends them to the cloud service over HTTP using the wireless communication system. Figure 4 shows sample formatted data that are sent to the cloud service for analyzing the vehicle's current situation.

*3.2.2. 3G Module.* The 3G module provides the wireless communication facility to the control unit. Cell phone operators offer 3G network coverage that provides the fastest Internet connections over a mobile network in most of the developing countries so far. The amount of data and the frequency of sending the data from the water-based vehicle to the cloud service demand this type of network connection. 3G modules like SIM5216E ensure wireless communication in case of unavailability of 3G network coverage [20].

### 3.3. GPS System.
It is mainly responsible for determining the detailed information regarding the vehicle location, including latitude and longitude, time, and speed [21] and sending the information to the collector so that the vehicle's location and distance from the different rescue station can be determined in any alarming situation.

### 3.4. Voice Communication.
This system enables the control room and the rescue teams to establish a direct voice connection to a vehicle when necessary. The master of the vehicle can also establish a direct voice connection between the driving room and the control room. For example, the control room and the rescue teams need to communicate with the vehicle driving room after they receive an alert or emergency signal from the application service. Apart from the monitors, voice communication system acts as a vital component of IWTS, especially during emergency.

### 3.5. Cloud Service.
The cloud service has two components, which are the application service and the database service. These services are deployed in the cloud as it has been observed that cloud computing helps in reducing the M2M implementation cost as the complex data processing held on the server, which also enhances the performance of the M2M devices [26]. Furthermore, Ajah et al. stated that this type of

```
data = {
    gps: {ip: "82.62.10.1", lat: "12543.5", long: "123456.60", speed: 20.00, time: "16:45"},
    passenger: {ip: "82.62.10.2", count: 300, weight: "20000"},
    air: {ip: "82.62.10.3", weather: "foggy", direction: "north-south", speed: 100},
    wave: {ip: "82.62.10.4", height: 100, direction: "north-south", speed: 45},
    floating-point: {ip: "10.62.10.5", ref-value: 100, actual-value: 101},
    .......
}
```

Figure 4: Sample JSON formatted data that are sent to the cloud service from the vehicle.

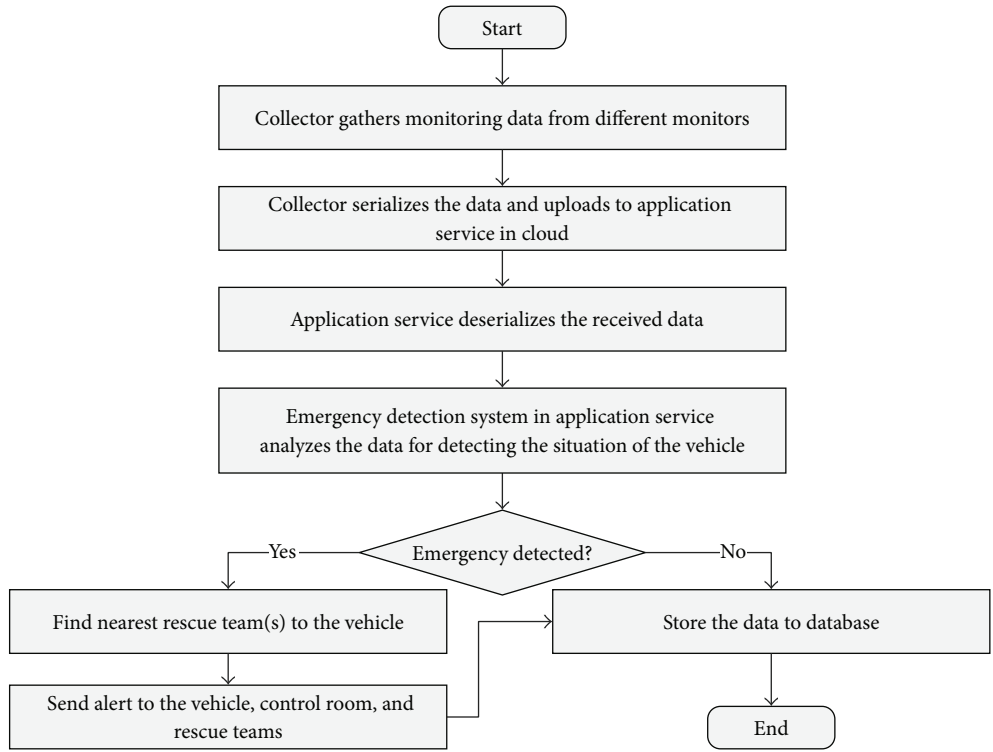

Figure 5: Decision-making and monitoring flow chart for vehicle situation analysis in IWTS.

centralized data processing mechanism (cloud computing) is employed to enhance the energy efficiency of M2M devices as the sensor based M2M devices sense information and do a very little computation, and the complex computation of data takes place in the central application server [27]. GPRS technology is used to send notifications to the moving vehicle and other systems [28].

*3.5.1. Application Service.* After receiving the data from the control unit of the vehicle, it processes the data and analyzes the data with the help of the emergency prediction system built and trained with the help of different data mining algorithms (e.g., data fusion algorithms [29, 30], Bayesian Belief Network [31]). After analyzing the data, it sends the necessary information to the transportation authority and the moving vehicle's control unit and to the station terminal using the Internet and GPRS technology. Figure 5 shows how the monitoring data flows and vehicle situation is analyzed in IWTS. Data pulling from the cloud services at any time through the Internet is considered as a regular feature. If application service predicts any emergency or unusual situation after analyzing the data, it sends an alarm to the transportation authority, vehicle's control unit, and station terminal as well as to the nearest rescue station for initiating communication to the vehicle and rescue operation. The information includes vehicle information, the location, and the predicted situation with other necessary details.

To develop the decision model to predict emergencies and unusual situations with the sensor data sent from the control unit of the running vehicle, data fusion concept can be used. Faouzia et al. provide a survey of how data fusion techniques are used in distinctive areas of ITS and how incidents are detected [29]. Moreover, Leppänen et al. develop a service using artificial intelligence where they use different data fusion algorithms for analyzing sensor data and taking decisions in ITS [30]. Hence, these techniques might be the recommendation for now as they are also applicable for IWTS. A recent research conducted by Khaled and Kawamura has shown the probability analysis of collusion between vessels in Chittagong port using Bayesian Belief Network [31]. For their research, they analyzed the accident database of Chittagong port. Although such accident database is hard to realize for the other waterways for transporting people, their approach might be a potential starting point for developing the emergency prediction system for IWTS. Figure 6 presents the cloud based process flow model for emergency detection for water-based vehicles in IWTS.

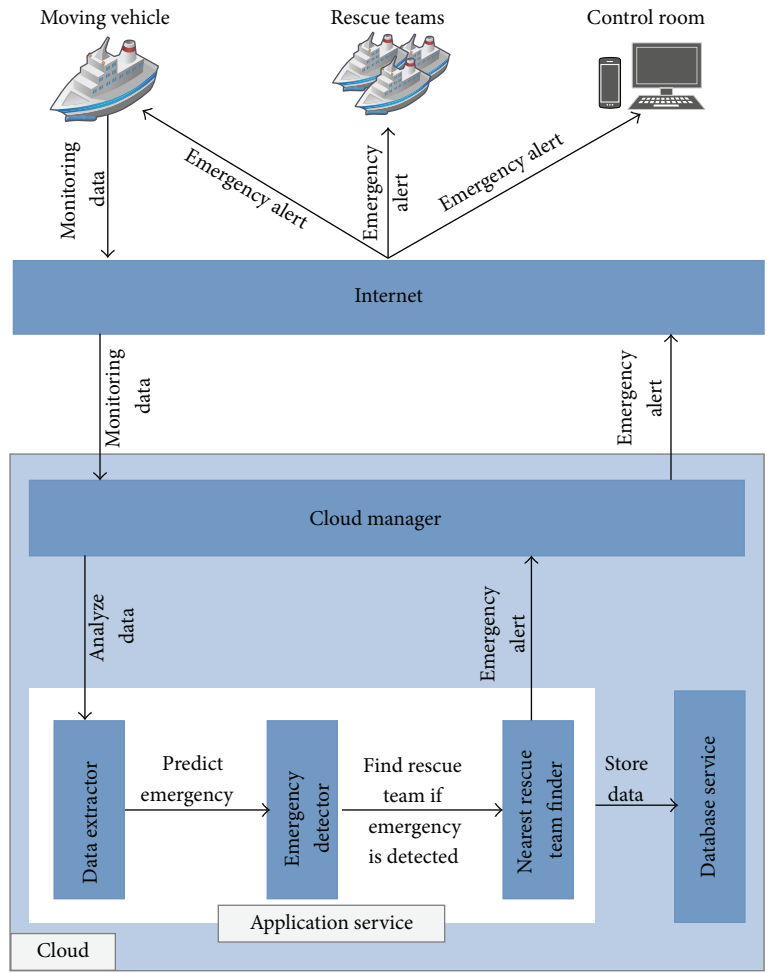

Figure 6: Cloud based process flow model for vehicle situation analysis in IWTS.

*3.5.2. Database Service.* The application service stores the data it receives from the running vehicles and other sources without any data reduction for future use and to maintain historical information for each and every event, which enables IWTS to be connected to other applications and services in the future.

## 4. Impacts of IWTS on Safety in Waterways

With IWTS, every passenger vehicle in water is under IWTS's network coverage. The monitoring system of IWTS installed in each passenger vehicle sends monitoring data to the application service at a particular interval which ensures that a vehicle's status in water is always known. Thus IWTS has inevitable impacts on the improvement in the safety scenario in the waterways, and some of them are mentioned below:

(i) Vehicles in water are constantly under monitoring as various monitoring devices are monitoring the vehicle and its circumstances, and the information is sent regularly to the cloud service for observation.

(ii) The vehicle overweight condition is constantly monitored as the passenger counter and the weight monitor let the authority know about the number of passengers and the total weight taken on board.

(iii) Since the vehicle's status is always known, illegal passenger overloading will be reported immediately and as other natural issues are consistently taken care of proactively, the possibility of occurrence of accidents will be reduced drastically.

(iv) As the information about any emergency and accidents is predicted immediately, the rescue operations become easier and faster. This brings a radical decrease in the loss of lives and properties.

(v) Since the vehicle location is always known with its status and the information is available through the Internet, passengers may have the opportunity to have real-time forecasting of their journey.

(vi) It enhances the safety of the water transportation system, which leads to an increased use of this mode of transportation.

(vii) Although the number of people using this route increases, the station management becomes easier.

(viii) All of the above positives will have a positive impact on the country's economy as well.

## 5. Experiment Result

The primary focus of this research is to be able to detect any emergency situation of vehicles in waterways and send emergency alerts to the monitoring authority for initiating an action to stop any mishaps in waterways. Although the proposed model promises to monitor a number of different areas, data from only two areas were used in this experiment to detect any emergency. The chosen areas were passenger weight monitoring and freeboard (the height of the water-based vehicle's deck above the water level) monitoring. This experiment was conducted in a live Virtual Private Server (VPS) configured with Intel® Xeon® CPU E5-2630L 0 @ 2.00 GHz 64-bit operating system with 8.0 GB RAM installed in it. To examine the process of emergency detection and sending emergency alert to the control room and the rescue teams, Microsoft.Net development platform was used.

A scenario for passenger weight and freeboard monitoring of a water-based vehicle was assumed to determine a set of synthetic data for this experiment. According to the assumed scenario, the vehicle starts its regular journey from Barisal to Dhaka at 11:05 AM every day. Before starting its journey, it loads passengers from 10:00 AM to 11:00 AM. The maximum allowed weight that the vessel could carry was set to 14000 kgs and the freeboard was set to be at least 150 cm. The monitoring system of the vehicle was assumed to be sending data to the cloud service in an interval of 5 minutes. However, when the maximum allowed weight level was crossed and the freeboard started to decrease, the monitoring device started sending data to the cloud service more frequently (e.g., every one-minute interval). The synthetic dataset was generated for the passenger loading period (10:00 AM to 11:00 AM) only for seven consecutive days. To understand the characteristics of the system, this dataset was used to predict the event that crossed the threshold and identify the timing of the event. The program that was used for the experiment was written using C# programming language. Figure 7 represents the characteristics of the synthetic data for weight monitoring, and Figure 8 represents the characteristics of the synthetic data for freeboard monitoring.

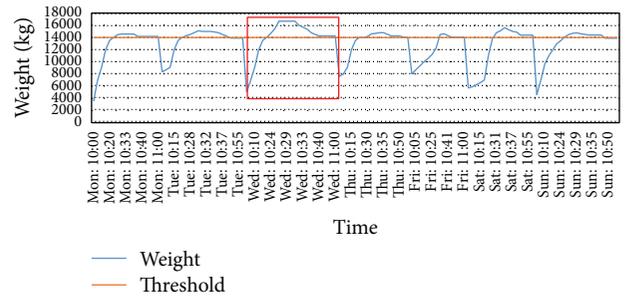

Figure 7: Characteristics of synthetic data representing the weight monitoring for seven consecutive days of the IWTS.

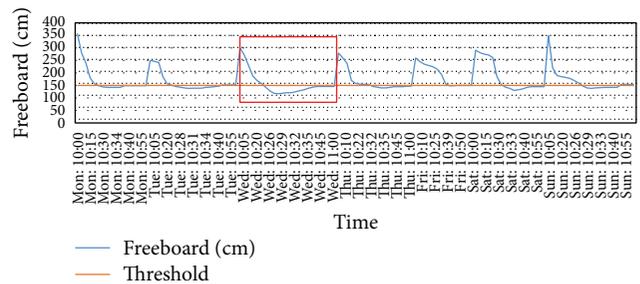

Figure 8: Characteristics of synthetic data representing the freeboard monitoring for seven consecutive days of the IWTS.

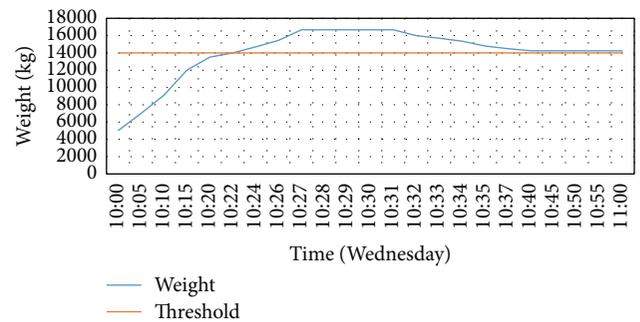

Figure 9: Zoomed view of the marked area of Figure 7 that represents the characteristics of the synthetic data for the weight monitoring on Wednesday.

The charts provided in Figures 7 and 8 show that, among the seven days, there was an increasing peak in weight loaded into the vehicle on Wednesday that forced the freeboard to go down gradually as well. The peaks are marked with a red rectangle; Figures 9 and 10 represent the zoomed view of those marked peaks of Figures 7 and 8.

From the charts provided in Figures 9 and 10, it can be seen that around 10:22 AM on Wednesday the maximum allowed weight level or threshold was crossed, and the freeboard started to decrease. Around 10:28 AM on the same day, the weight reached its peak while the freeboard was the lowest. At that point, high level emergency alert was sent to the control room and the nearest rescue team. Upon receiving the alert, the control room and the nearest rescue team looked into the matter and the vehicle got back to its normal condition by unloading extra passengers at around 10:40 AM.

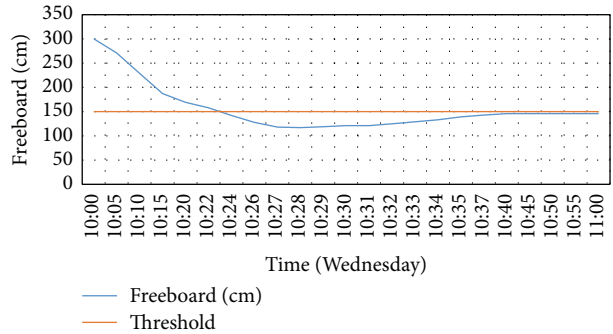

Figure 10: Zoomed view of the marked area of Figure 8 that represents the characteristics of the synthetic data of the freeboard monitoring on Wednesday.

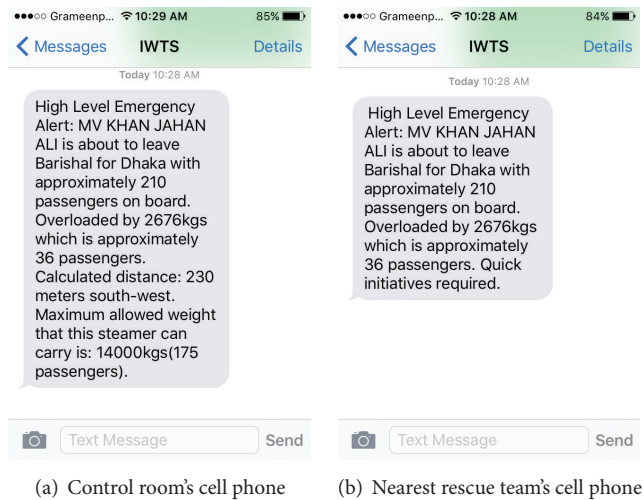

(a) Control room's cell phone

(b) Nearest rescue team's cell phone

Figure 11: Emergency alert message to the control room and the nearest rescue team.

A simple C# program was written to detect emergency event from the passenger weight and freeboard monitoring data. The application was hosted in the VPS and the data was sent from a client computer to the application over http using a 5 mbps shared broadband Internet connection. Upon receiving the data, the server program processed it and generated emergency alert when the passenger weight went beyond the threshold or the freeboard was below the threshold. Hence, when the passenger weight started going above the threshold and freeboard was going down the threshold, respectively, the application started sending emergency alerts to the control room and the nearest rescue teams using GPRS. Figures 9 and 10 show that around 10:28 AM the weight reached its peak, and the freeboard was the lowest, and a high level emergency alert was generated by the server program. Figure 11 represents the high level emergency alert messages sent to the cell phone of the control room and the nearest rescue team. According to the experiment, the entire process took about 2 to 3 seconds.

Although the above experiment was performed with only two types of data, it clearly shows that if passenger overloading can be stopped, it drastically reduces the probability of mishaps in waterways as passenger overloading is considered to be one of the most predominant reasons for mishaps in waterways. Thus it enhances the safety of the water-based vehicles significantly. Moreover, when there will be other monitoring devices in place, the safety will be enhanced greatly and the whole waterways will turn into an intelligent and safe transportation system.

## 6. Future Scope of Work

Although this paper has considered almost all the aspects of the inland water transport system for the developing countries, there may remain some scope of improvement and enhancement of the proposed system. Further research is needed to develop the architecture of an efficient decision model of an emergency prediction system of application service of IWTS since there are challenges involved in gathering previous real data due to its unavailability to train the decision model. Moreover, automating the vehicles used by the rescue team and connecting those vehicles to IWTS might be another good candidate for future research. Furthermore, achieving real-time forecasting of passenger journey from the data available in IWTS can be considered as another candidate for further research.

## 7. Conclusion

As it is observed that there are frequent unpleasant incidents occurring in the waterways which cause loss of human lives as well as causing enormous loss to developing countries like Bangladesh, it is very important to have a safe and efficient water transportation system for these countries. The proposed model makes the system smart and safe for the people. The system can easily detect any alarming situation for the vehicle moving in the water. Thus the situations can be proactively handled and in case of emergency the rescue operation is initiated immediately, which will save lives and make the transport system safe and reliable. Although this model is developed with a focus on Bangladesh, it can be fitted easily to other developing countries with minimal change or even without any change as it is as the availability of resources, technologies, and economic capability of the developing countries are of similar types.

## Conflict of Interests

The authors declare that there is no conflict of interests regarding the publication of this paper.

## Acknowledgment

The authors extend their appreciation to the Deanship of Scientific Research at King Saud University, Riyadh, Saudi Arabia, for funding this work through Research Group Project no. RG-1436-016.

## References

[1] T. Kalyani, D. S. P. Vidyasagar, and V. S. J. Srinivas, "Accident analysis of river boats capsize in Indian inland waters and safety aspects related to passenger transportation," *International Journal of Innovative Research & Development*, vol. 4, no. 7, pp. 8–17, 2015.

[2] Z. I. Awal, M. R. Islam, and M. M. Hoque, "Marine vehicle accident characteristics in bangladesh: study on collision type accidents," in *Proceedings of the International Conference on Mechanical Engineering (ICME '07)*, Dhaka, Bangladesh, 2007.

[3] BIWTA, "About Us: Bangladesh Inland Water Transport Authority," BIWTA, 2014, http://www.biwta.gov.bd/website/?page_id=2.

[4] MMM Group, *Bangladesh: Capacity Building and Support to the Transport Sector Coordination Wing of the Planning Commission*, Planning Commission, Dhaka, Bangladesh, 2012.

[5] Z. I. Awal, "A study on inland water transport accidents in Bangladesh: experience of a decade (1995–2005)," in *Proceedings of the International Conference on Coastal Ships and Inland Waterways*, Royal Institution of Naval Architects, 2006.

[6] P. Karmakar, "Death toll reaches 70," February 2015, http://www.thedailystar.net/frontpage/death-toll-reaches-70-3656.

[7] Pinak-6 search abandoned, The Daily Star, August 2014, http://www.thedailystar.net/pinak-6-search-abandoned-36738.

[8] S. Mamun, "Despite 10,000 deaths, shipping safety beyond horizon," *DhakaTribune*, 2015, http://www.dhakatribune.com/bangladesh/2015/feb/24/despite-10000-deaths-shipping-safety-beyond-horizon.

[9] R. M. Cardoso, N. Mastelari, and M. F. Bassora, "Internet of things architecture in the context of intelligent transportation systems—a case study towards a web-based application deployment," in *Proceedings of the 22nd International Congress of Mechanical Engineering (COBEM '13)*, São Paulo, Brazil, 2013.

[10] L. Huang and C. Liu, "The application mode in urban transportation management based on internet of things," in *Proceedings of the 2nd International Conference on Computer Science and Electronics Engineering (ICCSEE '13)*, Atlantis Press, Hangzhou, China, March 2013.

[11] C. YuFeng, X. ZhengTao, and C. Li, "The research development on the transportation Information collection technology of Intelligent Transportation system," *Hubei Automobile Industry Institute Journal*, vol. 24, no. 2, pp. 30–36, 2010.

[12] E. Fleisch, "What is the internet of things? An economic perspective," *Economics, Management, and Financial Markets*, vol. 5, no. 2, pp. 125–157, 2010.

[13] G. Lawton, "Machine-to-machine technology gears up for growth," *Computer*, vol. 37, no. 9, pp. 12–15, 2004.

[14] C. Pereira and A. Aguiar, "Towards efficient mobile M2M communications: survey and open challenges," *Sensors*, vol. 14, no. 10, pp. 19582–19608, 2014.

[15] J. Wan, M. Chen, F. Xia, D. Li, and K. Zhou, "From machine-to-machine communications towards cyber-physical systems," *Computer Science and Information Systems*, vol. 10, no. 3, pp. 1105–1128, 2013.

[16] M. D. Ramos, A. D. Foster, S. Felici-Castell, V. G. Fos, and J. J. P. Solano, "Gatherer: an environmental monitoring application based on IPv6 using wireless sensor networks," *International Journal of Ad Hoc and Ubiquitous Computing*, vol. 13, no. 3-4, pp. 209–217, 2013.

[17] Z. M. Fadlullah, M. M. Fouda, N. Kato, A. Takeuchi, N. Iwasaki, and Y. Nozaki, "Toward intelligent machine-to-machine communications in smart grid," *IEEE Communications Magazine*, vol. 49, no. 4, pp. 60–65, 2011.

[18] J.-S. Lee, "Performance evaluation of IEEE 802.15.4 for low-rate wireless personal area networks," *IEEE Transactions on Consumer Electronics*, vol. 52, no. 3, pp. 742–749, 2006.

[19] J.-S. Lee, Y.-W. Su, and C.-C. Shen, "A comparative study of wireless protocols: bluetooth, UWB, ZigBee, and Wi-Fi," in *Proceedings of the 33rd Annual Conference of the IEEE Industrial Electronics Society (IECON '07)*, pp. 46–51, IEEE, Taipei, Taiwan, November 2007.

[20] Q. Zhang and Y. Chen, "Design of wireless intelligent video surveillance system based on 3G network," *TELKOMNIKA Indonesian Journal of Electrical Engineering*, vol. 12, no. 1, pp. 206–217, 2014.

[21] Z. Yongjun, Z. Xueli, Z. Shuxian, and G. shenghui, "Intelligent transportation system based on internet of things," in *Proceedings of the World Automation Congress (WAC '12)*, pp. 1–3, IEEE, Puerto Vallarta, Mexico, June 2012.

[22] Y. Wang and H. Qi, "Research of intelligent transportation system based on the internet of things frame," *Wireless Engineering and Technology*, vol. 3, no. 3, pp. 160–166, 2012.

[23] Stand-Alone Directional Passenger Counter, INFODEV, January 2015, http://www.infodev.ca/vehicles/products-and-passenger-counters/products/counting-devices/da-200.html.

[24] J. Matamoros and C. Antón-Haro, "Data aggregation schemes for Machine-to-Machine gateways: interplay with MAC protocols," in *Proceedings of the 21st Future Network & Mobile Summit (FutureNetw '12)*, pp. 1–8, IEEE, Berlin, Germany, July 2012.

[25] D. Boswarthick, O. Elloumi, and O. Hersent, Eds., *M2M Communications: A Systems Approach*, John Wiley & Sons, 2012.

[26] T. Osawa, "Practice of M2M connecting real-world things with cloud computing," *Fujitsu Scientific & Technical Journal*, vol. 47, no. 4, pp. 401–407, 2011.

[27] S. Ajah, A. Al-Sherbaz, S. Turner, and P. Picton, "Machine-to-machine communications energy efficiencies: the implications of different M2M communications specifications," *International Journal of Wireless and Mobile Computing*, vol. 8, no. 1, pp. 15–26, 2015.

[28] N. Chadil, A. Russameesawang, and P. Keeratiwintakorn, "Real-time tracking management system using GPS, GPRS and Google earth," in *Proceedings of the 5th International Conference on Electrical Engineering/Electronics, Computer, Telecommunications and Information Technology (ECTI-CON '08)*, pp. 393–396, Krabi, Thailand, May 2008.

[29] N.-E. El Faouzi, H. Leungd, and A. Kuriand, "Data fusion in intelligent transportation systems: progress and challenges—a survey," *Information Fusion*, vol. 12, no. 1, pp. 4–10, 2011.

[30] T. Leppänen, M. Perttunen, P. Kaipio, and J. Riekki, "Sensor data fusion middleware for cooperative traffic applications," *International Journal on Advances in Networks and Services*, vol. 4, no. 1-2, pp. 149–158, 2011.

[31] E. Khaled and Y. Kawamura, "Application of bayesian belief network to estimate causation probability of collision at chittagong port by analyzing accident database of Bangladesh," in *Proceedings of the Japan Society of Naval Architects and Ocean Engineers (JASNAOE '14)*, Sendai, Japan, 2014.

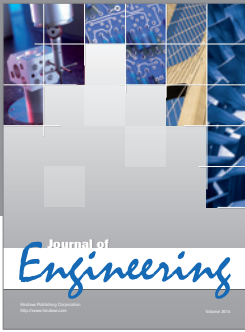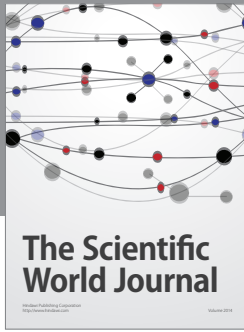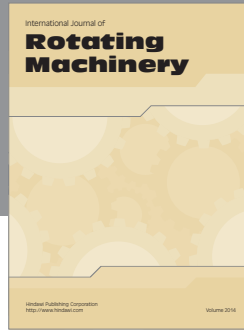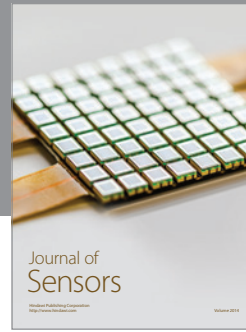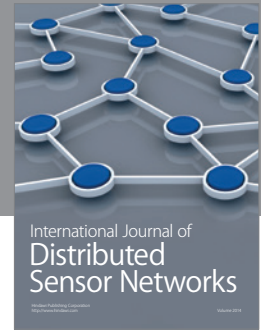
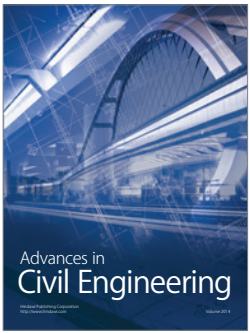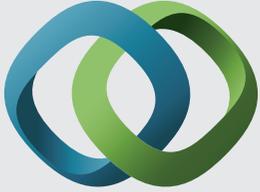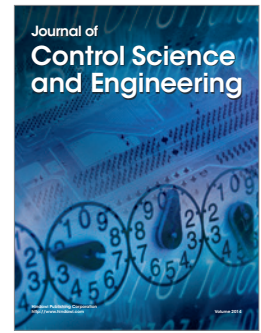
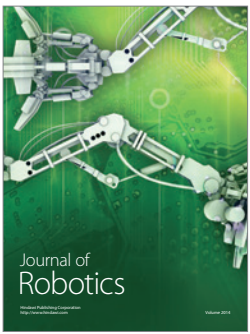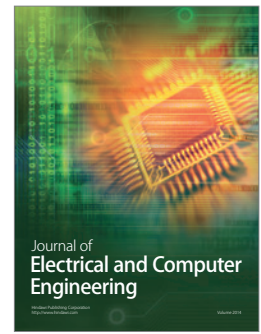
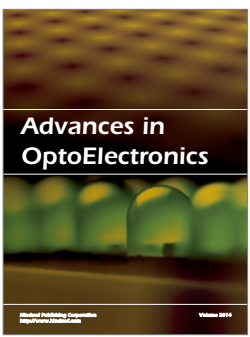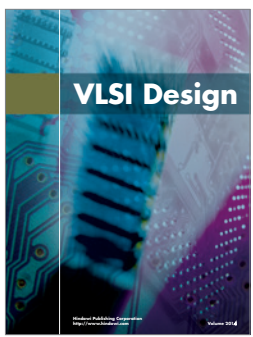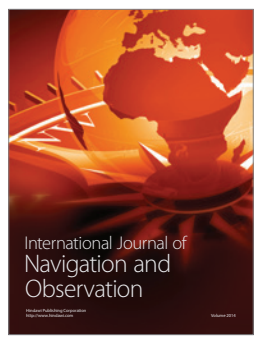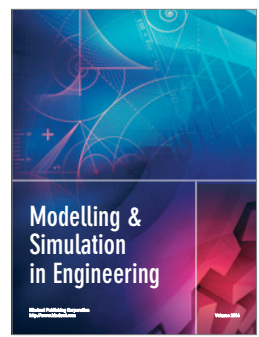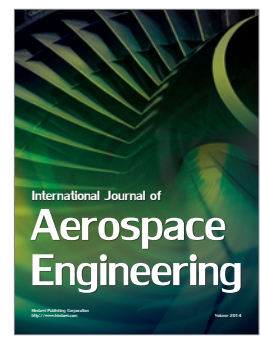
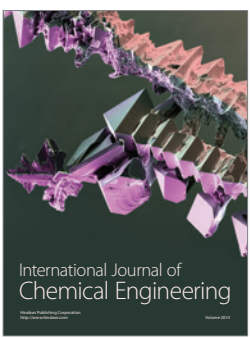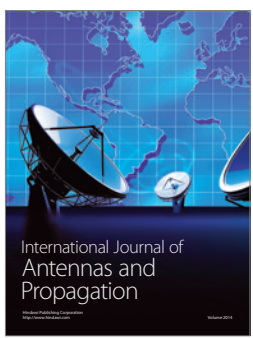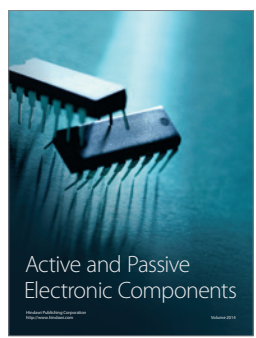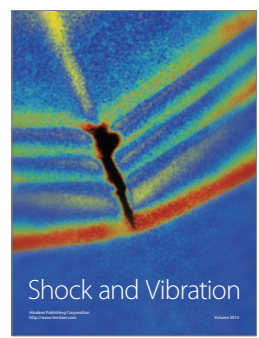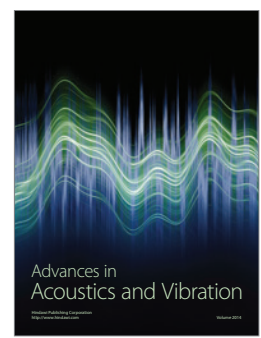